\documentclass{aa}
\usepackage{upgreek}
\usepackage{graphicx}
\usepackage[varg]{txfonts}

\usepackage{natbib}

\usepackage{soul}
\newlength{\linwx}
\usepackage{xcolor}
\usepackage{hyperref}
\usepackage{array}
\usepackage{amsmath}
\usepackage{multirow}
\usepackage{booktabs}

\hypersetup{colorlinks=true,linkcolor=purple,urlcolor=purple,citecolor = teal}

\graphicspath{ {Images/} }

\newcommand{\be}{\begin{equation}}
\newcommand{\ee}{\end{equation}}

\title{How to make giant planets via pebble accretion}

\author{Sofia Savvidou
\and Bertram Bitsch}

\offprints{S. Savvidou,\\ \email{savvidou@mpia.de}}
\institute{
Max-Planck-Institut f\"ur Astronomie, K\"onigstuhl 17, 69117 Heidelberg, Germany
}
\date{Received date / Accepted date }

\begin{document}

\abstract{Planet formation is directly linked to the birthing environment that protoplanetary disks provide. The disk properties determine whether a giant planet will form and how it evolves. The number of exoplanet and disk observations is consistently rising, however, it is not yet possible to directly link these two populations. Therefore, a deep theoretical understanding of how planets form is crucial. Giant planets are not the most common exoplanets, but their presence in a disk can have significant consequences for the evolution of the disk itself and the planetary system undergoing formation. Their presence also offers more chances of spotting observational features in the disk structure.
We performed numerical simulations of planet formation via pebble and gas accretion, while including migration, in a viscously evolving protoplanetary disk, with dust growing, drifting, and evaporating at the ice lines. In our investigation of the most favorable conditions for giant planet formation, we find that these are high disk masses, early formation, and a large enough disk to host a long-lasting pebble flux, so that efficient core growth can take place before the pebble flux decays over time. Specifically, core growth needs to start before 0.9 Myr to form a giant, with an initial disk mass of 0.04 $M_{\odot}$ (or higher) and the disk radius needs to be larger than 50 AU. However, small disks with the same mass allow more efficient gas accretion onto already formed planetary cores, leading to more massive gas giants. Given the right conditions, high viscosity ($\alpha=10^{-3}$) leads to more massive cores (compared to $\alpha=10^{-4}$) and it also enhances gas accretion. At the same time, it causes faster type II migration rates, so the giants have a decreasing final position for increasing viscosity. Intermediate dust fragmentation velocities, between 4 and 7 m/s, provide the necessary pebble sizes and radial drift velocities for maximized pebble accretion with optimal pebble flux. The starting location of a planetary embryo defines whether a giant planet will form, with the highest fraction of giants originating between 5 and 25 AU. Finally, a dust-to-gas ratio of 0.03 can compensate for lower disk masses with $f_{DG}\leq0.015$, but early formation is still important in order to form giant planets. We conclude that there is no specific initial parameter that leads to giant planet formation; rather, it is the outcome of a combination of complementary factors. This also implies that the diversity of the exoplanet systems is the product of the intrinsic diversity of the protoplanetary disks and it is crucial to take advantage of the increasing number and quality of observations to constrain the disk population properties and ultimately devise planet formation theories.}


\maketitle

\section{Introduction}

Protoplanetary disks provide the materials and environmental conditions for forming planets. This was recently reinforced by direct images of planets forming inside their natal environment \citep{Keppler+2018,Mueller+2018,Haffert+2019}. The disk properties could determine whether a planetary system contains a giant planet or not. The formation of a giant planet (or multiple ones) could subsequently influence the evolution of the planetary system and that of the disk itself, for example by influencing the dust mass evolution \citep{vanderMarelMulders2021}. 
At the same time, the presence of a giant planet offers more chances for observable features in protoplanetary disks, such as rings-gaps, spirals, or arcs \citep{Andrews2020}, which might not even coincide with the location of the planet(s) \citep{Bae+2018,Zhang+2018,Bergez-Casalou+2022}. For this reason, several studies have aimed to link giant planet
formation and evolution simulations with such features \citep[e.g.,][]{Pinilla+2012b,Dipierro+2015,Dong+2015a,Dong+2015b,Jin+2016,Fedele+2017,Teague+2018,Pinte+2018,Zhang+2018,Liu+2019,Muley+2019,Pinte+2020,Eriksson+2020,vanderMarel+2021}. 

Current exoplanet detection methods favor large planets at small distances from their host star, however, the ever-increasing number of detections allows for a statistical analysis of the exoplanet population, as long as each survey is combined with proper detection efficiency and completeness estimates \citep[see][for a review]{ZhuDong2021, WinnFabrycky2015}. Through the observations, we can estimate the occurrence fractions of planets, meaning the number of planets per star for a specific type of planet (e.g., divided by the planet's mass, radius, or orbital period) \citep{Wright+2012,Petigura+2013,Suzuki+2016,Mulders+2018}. This can then be converted to the fraction of stars hosting the specific type of planet. A combination of the detection surveys results in occurrence fractions of 10-20\% for all giant planets and specifically $\lesssim$ 16\% for warm and cold giants with orbits beyond 10 days \citep{Cumming+2008,Mayor+2011,Fernandes+2019,Fulton+2021,Rosenthal+2022}. 

Extracting information about the natal protoplanetary disks from exoplanet observations is almost unfeasible, even though the atmospheric abundances and the elemental ratios (such as C/O) could shed light on the formation pathways \citep[e.g.,][]{Oeberg+2011,Ali-Dib+2014,Helling+2014,Thiabaud+2014,Marboeuf+2014,Molliere+2015,Mordasini+2016,Madhusudhan+2017,Cridland+2019,SchneiderBitsch2021a,Bitsch+2022}. However, definitively linking observed atmospheric abundances to planet formation remains challenging \citep{Molliere+2022}. Thus, it is of utmost importance to properly understand the theoretical connection between the initial (disk) conditions and the possible planetary formation pathways.

We focus here on core formation via pebble accretion \citep{JohansenLacerda2010,OrmelKlahr2010,LambrechtsJohansen2012,Ormel+2017,JohansenLambrechts2017},
starting with an already formed planetary embryo and operating until its mass reaches the pebble isolation mass \citep{MorbidelliNesvorny2012,Lambrechts+2014,Ataiee+2018,Bitsch+2018}, which depends on the disk properties. 
In the classical core accretion scenario \citep{Pollack+1996}, when the core has acquired enough mass, gas accretion starts dominating the growth of the planet. The growing planet interacts gravitationally with the disk, pushing material away from its orbit, and in doing so, it opens a gap in the gas \citep[e.g.,][]{GoldreichTremaine1980,LinPapaloizou1986,Crida+2006,CridaMorbidelli2007}. Gas accretion itself can even help in deepening this gap \citep{CridaBitsch2017,Bergez-Casalou+2020,Ndugu+2021}. If this gap is deep enough, then a pressure trap is generated and blocks the dust from drifting inside the planet's orbit \citep{PaardekooperMellema2006}. However, even planets with 10\% the mass of Jupiter can create spiral shocks that decouple the larger dust particles from the gas and subsequently create a gap in the dust without forming a corresponding gap in the gas beforehand \citep{PaardekooperMellema2004,PaardekooperMellema2006}.

Planet formation via pebble accretion has been widely studied in recent years \citep[e.g.,][]{Lambrechts+2014,Levison+2015,Chambers2016,Morbidelli+2015,Bitsch+2015b, Bruegger+2018,Ndugu+2018,Bitsch2019,SavvidouBitsch2021,Bitsch+2021}
but with regard to giant planets, it has often focused on core growth, thus neglecting gas accretion \citep{LambrechtsJohansen2014,Andama+2022} or on specific examples \citep{JohansenLambrechts2017}, or they have investigated the influence of pebble flux calculations \citep{Bitsch+2019}, instead of including dust evolution self-consistently. In the aforementioned studies, some initial parameters were varied, such as the disk mass, dust-to-gas ratio, or viscosity, but the main goal was commonly to recreate the Solar System or the observed exoplanet populations.

In this work, we use a model that includes pebble and gas accretion, gap opening, and type-I/type-II migration as previous models \citep{Bitsch+2015b,Bitsch+2018,Voelkel+2022}, but also includes self-consistent grain growth, drift, and pebble evaporation at the ice lines \citep{SchneiderBitsch2021a}. Using this framework, we strive to generalize the results from previous studies through a parameter study and directly link the initial (disk) conditions to the formation of giant planets.  

This paper is structured as follows: In Sect. \ref{Sec:Model}, we present the model we used and the quantities examined in the parameter study. We present individual example cases with varying initial parameters in Sect. \ref{Sec:Growthtracks} and discuss our findings from the parameter study for the most favorable conditions for giant planet formation in Sect. \ref{Sec:Giant planet recipes}. We discuss the conclusions from our results and how they connect to previous studies in Sect. \ref{Sec:Discussion} and we summarize our conclusions in Sect. \ref{Sec:Summary}. 

\section{Methods}
\label{Sec:Model}

\subsection{Model description}

\begin{table}[]
\centering
\begin{tabular}{@{}ccc@{}}
\toprule[1.2pt]
Parameter               & Values               &                              \\ \midrule
$M_0$ {[}$M_{\odot}${]}  & 0.01, 0.04, 0.07, {\bf 0.1}   & initial disk mass            \\ \addlinespace
$R_0$ {[}$R_{\odot}${]}  & 50, 100, 150, {\bf 200}   & initial disk radius          \\ \addlinespace
$\alpha$                & {\bf 0.0001}, 0.0005, 0.001 & $\alpha$-viscosity parameter \\ \addlinespace
$t_0$ {[}Myr{]}          & {\bf 0.1}, 0.5, 0.9, 1.3   & starting time of embryo      \\ \addlinespace
$\alpha_{p,0}$ {[}AU{]} & 1-50 every 1  & initial position of embryo   \\ \addlinespace
$u_{frag}$ {[}m/s{]}    & 1, 4, 7, {\bf 10}    & fragmentation velocity     \\ \addlinespace
$f_{DG}$                & 0.01, {\bf 0.015}, 0.03 & dust-to-gas ratio    \\ \bottomrule[1.2pt]
\end{tabular}
\caption{Parameters used in the simulations. We mark in bold the standard set, which is used as a reference in Figs. \ref{Fig:Growthtracks_main}-\ref{Fig:Growthtracks_moret0moredtg}.}
\label{Tab:parameters}
\end{table}

We used the \texttt{chemcomp} code \citep{SchneiderBitsch2021a} which simulates planet formation via pebble and gas accretion and migration in a semi-analytical 1D model of a viscously evolving protoplanetary disk, while taking into consideration dust growth, drift, and evaporation and condensation at the evaporation fronts, for multiple chemical species. The following prescriptions are listed in more detail in \citealt{SchneiderBitsch2021a}. The viscosity in the model follows an $\alpha$-prescription \citep{ShakuraSunyaev1973} and we chose to keep the vertical stirring-settling parameter always constant to $a_z=10^{-4}$, motivated by the study of \citealt{Pinilla+2021}. These authors suggest that a low vertical mixing, which is different from the parameters controlling the turbulent velocities, the radial diffusion, and the gas viscous evolution, leads to better agreement with observations.

The midplane temperature is set by considering the viscous heating and the irradiation from the star. The midplane temperature sets the aspect ratio of the disk, which, in turn, determines the pebble isolation mass. However, here we kept the temperature fixed in time for simplicity, unlike our previous work \citep{Savvidou+2020}, where we linked the disk thermodynamics self-consistently with the dust sizes. The fixed temperature is motivated by the fact that the decrease in the temperature due to a gradual decrease in the gas surface density was found to be minimal (see Appendix B in \citealt{SchneiderBitsch2021a}). Additionally, the self-consistent treatment followed in \citealt{Savvidou+2020} is computationally very expensive and it would still lead to minimal differences (especially with the lower viscosity used here) in the general trends that we aim to discuss in this work.

The dust growth was modeled using the two populations approach described in \citealt{Birnstiel+2012}, which means that we take growth, fragmentation, and drift, along with drift-induced fragmentation into account. Thus, we assume spherical compact dust grains, while neglecting the effects of porosity, which would allow for compaction and for smaller grain sizes to bear the maximum mass values \citep{Okuzumi+2012,Estrada+2022}\footnote{ Such fractal aggregates could lead to smaller Stokes numbers in the outer regions, thus retaining significant mass there longer and aiding giant planet formation. However, here, we generalize the effects of the dust distribution through the fragmentation velocity and the influence of viscosity on the maximum grain size (see Sect. \ref{Sec:Initial_conditions}).}. This also means that the dust surface density, used in the pebble accretion prescription, is a product of the disk evolution itself. This is similar to our previous work \citep{SavvidouBitsch2021} and others \citep[e.g.,][]{Venturini+2020,Drazkowska+2021}, but lies in contrast to more simplified approaches \citep[e.g.,][]{LambrechtsJohansen2014,Bitsch+2015b,Ndugu+2018}. 

In this code, planet formation begins by inserting planetary embryos in the disk with the pebble transition mass as their initial mass, which is expressed as: 
\be \label{Eq:M_t} M_{\rm{t}}=\sqrt{\frac{1}{3}}\frac{{\Delta v}^3}{G\Omega}~,\ee
and is relevant to protoplanets accreting pebbles in the Hill (shear) regime \citep{LambrechtsJohansen2012,JohansenLambrechts2017}. At this mass, pebble accretion starts becoming efficient. Therefore, the starting mass for each depends on the local disk conditions, $\Delta v$, the sub-Keplerian speed of the particles, and $\Omega$, the Keplerian angular frequency, with G being the gravitational constant.

The pebble accretion rates are given by \citep{JohansenLambrechts2017}:
\be \dot{M}_{\rm{peb}}=
\begin{cases}
2R_{\rm{acc}}\Sigma_{\rm{peb}}\delta v~,~\text{2D accretion,} \\
\pi R^2_{\rm{acc}}\rho_{\rm{peb}}\delta v~,~\text{3D accretion,} 
\end{cases}
\ee
with $R_{\rm{acc}}$ as the accretion radius of the growing protoplanet, $\Sigma_p$, $\rho_p$ as the pebble column density and midplane density, respectively, and $\delta v \equiv \Delta v +\Omega R_{\rm{acc}}$ as the approach speed, which (along with the column or midplane density) determines the azimuthal pebble flux. 
The radius of the accretion sphere (in the Hill limit, where $\Delta v \ll\Omega R_{\rm{H}}$) is:
\be R_{\rm{acc}} \propto \left(\frac{\Omega \rm{St}}{0.1}\right)^{1/3} R_{\rm{H}}~, \ee
where St is the pebble Stokes number and $R_{\rm{H}}$ the Hill radius of the embryo.
The transition from the 3D to the 2D regime happens when 
\be H_{\rm{peb}} < \frac{2\sqrt{2\pi}R_{\rm{acc}}}{\pi}~, \ee
following \citet{Morbidelli+2015}. We used the pebble scale height, $H_{\rm{peb}} = H_{\rm{gas}} \sqrt{\alpha_z/\rm{St}}$ and $\rho_{\rm{peb}} = \Sigma_{\rm{peb}}/(\sqrt{2\pi}H_{\rm{peb}})$.

The planetary embryos are inserted at a given time and the core starts growing with a 90\% contribution of the pebble accretion rate. The remaining 10\% contributes to the primary envelope to account for pebble evaporation during the core buildup in a simplified way. For the pebble isolation mass, we use
\be \label{Eq:pebiso} M_{iso}=25 f_{\rm fit} M_{\oplus}~, \ee
with
\be f_{\rm fit} = \left[\frac{H/r}{0.05}\right]^3 \left[0.34\left(\frac{\log(0.001)}{\log(\alpha)}\right)^4+0.66\right]~, \ee
adapted from \citealt{Bitsch+2018} without the dependence on the radial pressure gradient that is not very strong and not well studied around icelines.
When this limiting mass is reached, pebble accretion ends and the core has reached its final mass \citep{MorbidelliNesvorny2012,Lambrechts+2014,Bitsch+2018,Ataiee+2018}.
The envelope then quickly contracts and starts accreting gas. The approach followed in this model is the same as in \citealt{Ndugu+2021}, where the gas accretion rate is given by the minimum between the accretion rates given by \citealt{Ikoma+2000} and \citealt{Machida+2010}, and by the gas that the disk can viscously provide into the horseshoe region after the planet has emptied it \citep{Ndugu+2021}.

The forming planet migrates using the \citealt{Paardekooper+2011} type-I migration rates related to the Lindblad, the barotropic, and the entropy-related corotation torques. The code also includes the effects of the thermal \citep{Masset2017} and the dynamical torques \citep{Paardekooper2014}. If gas accretion becomes efficient, the planet can then open a gap, first via the gravitational interaction of the disk and the planet and then due to the gas accretion as in \citealt{Ndugu+2021}. If the planet carves a deep enough gap, it starts migrating in the type-II regime with a viscosity-dependent rate.  The chemical model and a more detailed description of the code can be found in \cite{SchneiderBitsch2021a}. 

\subsection{Initial conditions}
\label{Sec:Initial_conditions}

We perform a parameter study varying the disk mass ($M_{disk}$), disk radius ($R_{disk}$), $\alpha$-viscosity, the dust fragmentation velocity ($u_{frag}$), and, finally, the time ($t_0$) and location ($a_p$) of the inserted planetary embryo. The values chosen for each of those parameters are summarized in Table \ref{Tab:parameters}. The nominal dust-to-gas ratio is chosen as $f_{DG}=1.5\%$, but we also compare with $f_{DG}=1\%$ and $f_{DG}=3\%$. 
The maximum disk mass is chosen based on the maximum disk masses from the current disk mass estimations \citep{Manara+2022arXiv} and the minimum is based on a reasonable value for solar-type stars, keeping in mind that our goal is to form giant planets. We also choose the disk radii range to cover the most commonly observed ranges \citep[e.g.,][]{Pascucci+2016,Ansdell+2016,Ansdell+2017,Long+2019,Maury+2019,Ansdell+2020,Tobin+2020,Sheehan+2020,Sanchis+2021}. 

We already expect that early formation will be helpful for giant planet formation, so we choose all planetary embryos to be inserted before 1.3 Myr and we position them from 1 to 50 AU every 1 AU. 
It should be noted that we always simulate the growth of one planet at a time for each disk, therefore the interactions between multiple planets are neglected. However, planet formation by pebble accretion is more efficient in the inner disk regions, meaning that if giant planets form in the outer disk, they would also form in the inner disk. At the same time, if the pebble flux allows for the formation of one giant, then multiple giants can also form because planets growing by pebble accretion only reduce the pebble flux by a few percent \citep{Lambrechts+2014,Bitsch+2019,Matsumura+2021}. 

The star in the simulations is of solar mass. The integration stops at around 3 Myr, assuming that at this point, photoevaporation significantly depletes the disk from the gas \citep{Mamajek2009}. Specifically, at 3 Myr, we introduce an exponential decay of the surface density of the disk, reducing it to 0 within 100 kyr. If a planet migrates to the inner edge of the disk, we stop the accretion onto the planet. This is, in part, motivated by recycling flows that can prevent the accretion of atmospheres at close distances \citep[e.g.,][]{Lambrechts+2017,Cimerman+2017,Moldenhauer+2021}. However, we let the simulation run until the end of the disk's lifetime to trace the dust evolution that follows, which we will discuss in future work. In addition, approaching this inner edge does not necessarily mean that the planet will be lost to the star because there could be a migration trap near the inner rim \citep{Masset+2006,Romanova+2019,Flock+2019,AtaieeKley2021,Chrenko+2022}. This migration trap would probably be mainly efficient for low-mass planets, however, we chose to keep the planets in the disk, so that we track how the disk would evolve with the presence of a planet. Given the uncertainty of their fate, we also chose to keep them in our sample. 

The fragmentation velocity is the relative threshold velocity at which dust growth is halted and particles fragment instead. Within the two populations model, \citep{Birnstiel+2012} the maximum dust size before fragmentation is defined as
\be \label{Eq:s_max} s_{max} = f_f\frac{2\Sigma_g}{3 \pi \alpha \rho_s}\frac{u_{frag}^2}{c_s^2}~, \ee 
where $f_f$=0.37 is a best-fit parameter to the full grain growth model, $\rho_s$ is the dust grain density, and $c_s$ the local sound speed.
We chose to test a range from 1 to 10 m/s to cover different measurements from laboratory experiments \citep[e.g.,][]{Guettler+2010,GundlachBlum2015}. We kept the value constant in our model throughout the disk, motivated by laboratory experiments that show no difference in the sticking properties of water-ice and silicate aggregates \citep{MusiolikWurm2019,Steinpilz2019}. Our goal is to constrain giant planet formation via pebble accretion by examining which of the aforementioned initial conditions are the most favorable to form gas giants. 

\begin{figure}
\centering
\includegraphics[width=\columnwidth]{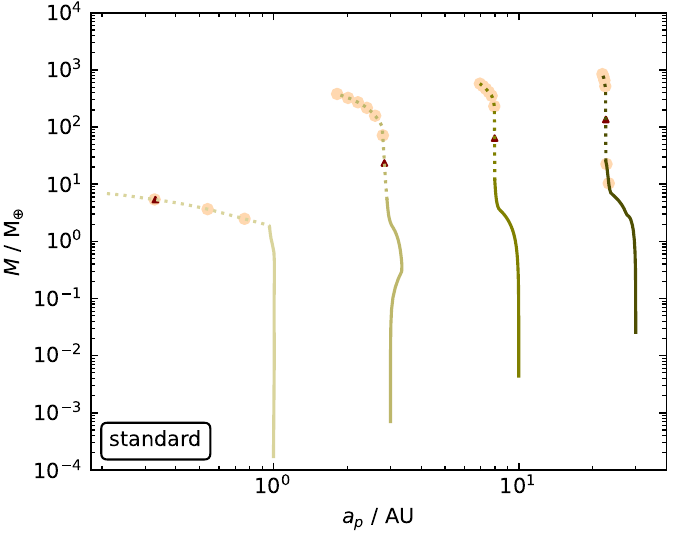}
\caption{Evolution of planetary mass as a function of position using the standard set of parameters (highlighted in bold in Table \ref{Tab:parameters}). The dots mark time steps of 0.5 Myr after the embryos start growing unless the planet has reached the inner edge, in which case they are omitted. The solid lines represent pebble accretion, while the dashed lines represent gas accretion. The triangles mark the transition from type-I to type-II migration.}
\label{Fig:Growthtracks_main}
\end{figure}

\begin{figure*}
\centering
\includegraphics[width=0.97\textwidth]{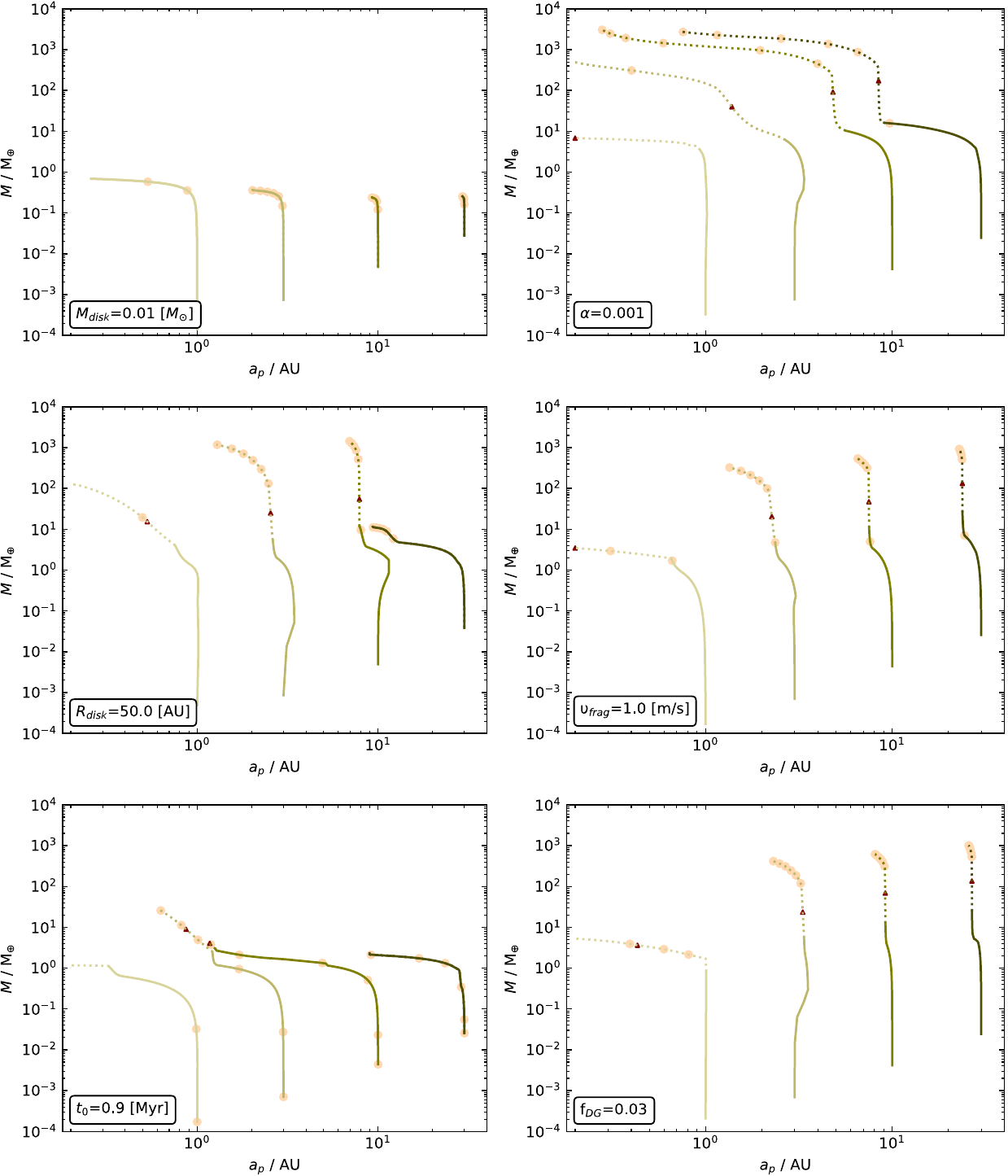}
\caption{Same as Fig. \ref{Fig:Growthtracks_main}, with one parameter changed (marked in each panel) compared to the standard model.}
\label{Fig:Growthtracks}
\end{figure*}

\begin{figure*}
\centering
\includegraphics[width=\textwidth]{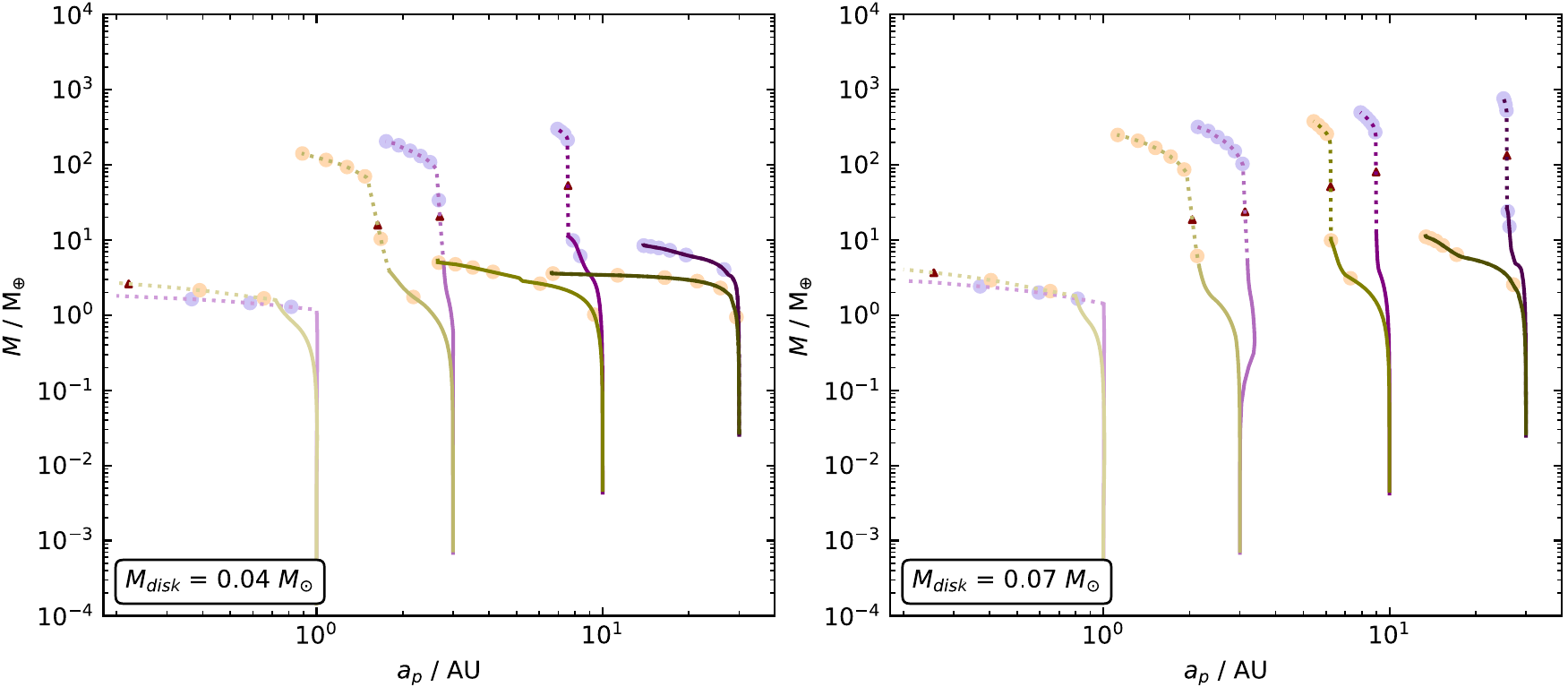}
\caption{Evolution of planetary mass as a function of position for two sets of simulations with $f_{DG}$=0.01 (green colors) and $f_{DG}$=0.03 (purple colors) and two different initial disk masses. The rest of the parameters are the same as in the standard case (bold in Table \ref{Tab:parameters}).}
\label{Fig:Growthtracks_moreM0moredtg}
\end{figure*}

\begin{figure*}
\centering
\includegraphics[width=\textwidth]{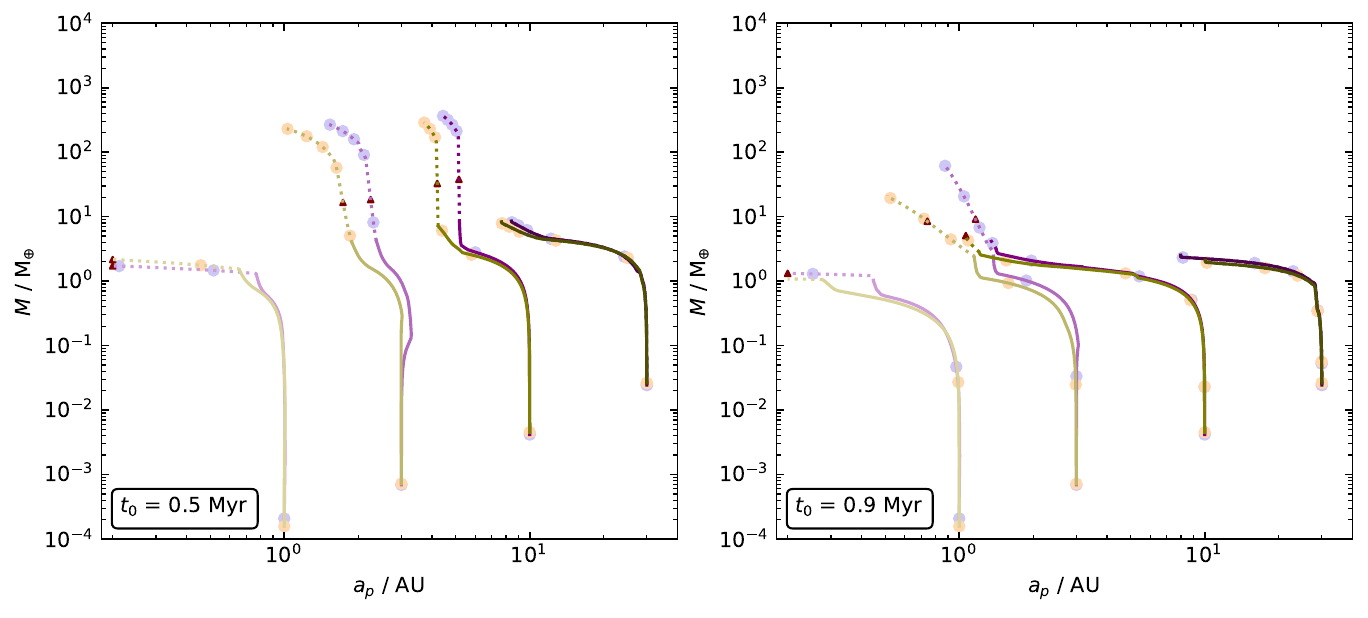}
\caption{Same as Fig. \ref{Fig:Growthtracks_moreM0moredtg}, for two different initial embryo injection times.}
\label{Fig:Growthtracks_moret0moredtg}
\end{figure*}

\section{Growth tracks}
\label{Sec:Growthtracks}

In this section, we discuss how the initial disk conditions influence the formation of giant planets through the growth tracks of specific example cases. First, we show in Fig. \ref{Fig:Growthtracks_main} the evolution of the planetary mass as a function of position for a specific set of parameters. The disk in this nominal case has the initial parameter values which are individually expected to be the most favorable for giant planet formation; high disk mass (0.1 $M_{\odot})$, large disk radius (200 AU), low viscosity ($10^{-4}$), early formation time (0.1 Myr), and high dust fragmentation velocity (10 m/s). These standard parameters are highlighted in bold in Table \ref{Tab:parameters}. The embryos of the examples shown here were placed at 1, 3, 10, and 30 AU. If the cores reach pebble isolation mass and the core mass has grown to several Earth masses, the envelope starts to slowly contract, followed by runaway gas accretion. The solid lines in Fig. \ref{Fig:Growthtracks_main} correspond to the core building phase, while the dashed lines correspond to the gas accretion phase. The dots mark timesteps of 0.5 Myr, counting from the time the embryo is inserted, but we omit them if the planet reaches the inner edge of the disk. 

Pebble accretion dominates over migration during the core buildup and the planets do not migrate inwards significantly. The planet starting at 3 AU also migrates slightly outwards within the first 0.5 Myr, as expected for low-mass planets around a few AU because of a positive heating torque \citep{BenitezLLambay2015,BaumannBitsch2020}. The low viscosity
leads also to a very slow type II migration during the gas accretion phase. However, the planet that starts at 1 AU has already reached the pebble isolation mass by 0.5 Myr; then, 1.5 Myr later, it reaches the inner edge of the disk. Having only a slow gas contraction phase, proportional to the core mass, it ends up being a 7 $M_{\oplus}$ sub-Neptune. For planets starting further out and this set of parameters, gas giants' formation is very efficient. There is enough material due to the initial high disk mass, the pebble flux can be sustained for a long time because of the large initial disk radius, and the low viscosity and large fragmentation velocity allow for large pebbles to form, enhancing the accretion rates. 

The pebble isolation mass increases as a function of orbital distance because of the increasing aspect ratio of the disk, owing to stellar irradiation (Eq. \ref{Eq:pebiso}), but most of the planets reach it before 0.5 Myr because of the abundant solid content of the disk. Nevertheless, it takes 1 Myr for the outermost planet ($a_p$=30 AU) to reach the pebble isolation mass because it is closer to the outer edge of the disk, where the surface density is decreased. In this case, in which the planets reach the pebble isolation mass, the higher core masses with increasing orbital distance lead to more efficient and timely gas accretion, therefore, the further out the embryo starts growing, the higher the final mass of the planet. Overall, the planets injected at 3, 10, and 30 AU grow to become cold gas giants with final masses of a few hundred $M_{\oplus}$.  

In Fig. \ref{Fig:Growthtracks}, only one initial disk parameter is changed in each plot. As expected, the initial disk mass is a very crucial parameter for giant planet formation. In the top, left plot the disk mass is reduced to 1\% of a solar mass and we see that at the end of the disk's lifetime, the planets are still in their core-building phase. The solid material is not enough for efficient core growth and the planets do not reach the pebble isolation mass before the disk disperses. All of their final masses are below 1 $M_{\oplus}$ and the innermost planet reaches the inner edge of the disk again, only this time after 1 Myr. 

If the initial disk radius is decreased (middle, left plot of Fig. \ref{Fig:Growthtracks}), then even the planet that started at 1 AU becomes a gas giant of around 100 $M_{\oplus}$ but it still migrates to the inner edge before 1 Myr. There is outward migration for both planets that start at 3 and 10 AU because of the heating torque. In this case, the disk radius is smaller but the total amount of material is the same in the disk, thus, locally, the dust and the gas surface density are higher, resulting in higher accretion rates. The heating torque is directly proportional to the solid accretion rate and since the latter is increased in this case, outward migration is more extended compared to the standard case in Fig. \ref{Fig:Growthtracks_main}. However, the closer the planets are to the outer edge, the more limited the available pebbles are. Especially in the case of a planet starting at 30 AU in this 50 AU disk, the pebble flux is significantly decreased and this leads to low accretion rates, so the planet does not reach the pebble isolation mass and, thus, it does not accrete gas, so it ends up as an ice giant of 11.5 $M_{\oplus}$. 

The time that the embryo has to grow is also a very crucial parameter.
In the bottom, left plot of Fig. \ref{Fig:Growthtracks}, we show the growth tracks for a starting time of 0.9 Myr instead of the nominal 0.1 Myr. We see that in this case only the planet starting from 3 AU accretes a gaseous envelope and its final mass is only a few Earth masses above Neptune's mass. If the injection time is delayed, then the dust has more time to radially drift inwards, so for the planets starting their growth at 10 And 30 AU, the pebble surface density has already decreased significantly, thus leading to slow pebble accretion rates.  The embryo starting at 1 AU, reaches the pebble isolation mass, as the ones starting at 3 and 10 AU, and it starts accreting gas but the low accretion rates mean that it migrates more and reaches the inner edge of the disk within 1 Myr. The embryo originating at 10 AU takes almost 3 Myr to reach the pebble isolation mass and for this reason, it only accretes around 1 $M_{\oplus}$ of gas.

 In the top-right plot the $\alpha$-viscosity parameter is instead $10^{-3}$. The higher viscous heating leads to higher pebble isolation masses (Eq. \ref{Eq:pebiso}) and it causes increased collisions, which keep the pebbles smaller (Eq. \ref{Eq:s_max}), leading to lower pebble accretion rates. Therefore, during the core-building phase, the planet accretes less material, in comparison to the model with $\alpha=10^{-4}$, and migrates inwards more. For this reason, the pebble isolation mass that the planet reaches is lower with $\alpha=10^{-3}$ and, thus, it is reached sooner, giving way to earlier gas accretion. At the same time, the gas accretion rate of the disk is larger due to the higher disk viscosity, resulting in more material delivered to the planetary horseshoe region, consequently increasing the planetary mass. The higher viscosity directly affects the type-II migration rates, so the planet migrates faster also during its gas accretion phase. The two planets that start growing closer to the star migrate to the inner edge of the disk before the end of the disk's lifetime. The planets that started at 10 and 30 AU end up as gas giants at 0.3 and 0.8 AU, respectively. The final masses are higher compared to the standard case, with the ones of the outermost gas giants being from a few hundred to around 3000 $M_{\oplus}$, compared to a few hundred in the standard case. 
 
The fragmentation velocity (middle-right plot of Fig. \ref{Fig:Growthtracks}) affects the final planetary masses only minimally in these examples. With this lower value, the pebble sizes remain small (Eq. \ref{Eq:s_max}), leading to slower pebble accretion; however, the radial drift velocities are also lower and the total pebble content remains enough for efficient and timely core growth. The final masses of the planets are slightly lower than the corresponding ones in the standard case of Fig. \ref{Fig:Growthtracks_main} because of the additional time it took for the planets to reach their pebble isolation masses and start accreting gas. The difference in the final positions is also minimal because migration is not directly affected by the different fragmentation velocity, it is only influenced by the competition with accretion.

 The final plot of Fig. \ref{Fig:Growthtracks} (bottom-right) shows the growth tracks of the four example planets with double the dust-to-gas ratio of the standard case shown in Fig. \ref{Fig:Growthtracks_main}. We find reduced migration because of the increased growth rates with the larger amounts of material available that cause a faster transition into the slower type-II migration, but the final masses remain almost the same. For the innermost planet, the pebble isolation mass is lower with the higher dust-to-gas ratio, even though at the beginning the higher dust-to-gas ratio leads to higher viscous heating because of the increased optical depth and the decrease in the gas surface density (check Eq. B.3 in \citealt{SchneiderBitsch2021a}) that both increase overall the temperature of the disk. The aspect ratio is then:
 \be H=\frac{c_s}{\Omega_K}~, \ee
 where $\Omega_K$ is the Keplerian angular velocity and the sound speed is: 
 \be c_s=\sqrt{\frac{k_BT_{mid}}{\mu m_p}}~ ,\ee 
 with $k_B$ as the Boltzmann constant, $T_{mid}$ as the midplane temperature, $\mu$ as the mean molecular weight, and $m_p$ as the proton mass. Therefore, the higher temperature leads to a higher aspect ratio and this directly influences the pebble isolation mass (Eq. \ref{Eq:pebiso}). However, the aspect ratio is actually evolving because of the evaporation of inward drifting pebbles that increases the vapor content of the gas, locally and over time (Eq. E.13 in \citealt{SchneiderBitsch2021a}). For this reason, in the disk with the higher dust-to-gas ratio, the gas gets more enriched and the mean molecular weight increases more over time, leading to a lower aspect ratio and, subsequently, a lower pebble isolation mass, mainly interior to the water ice line where the enrichment is significant. The innermost planet in the disk with $f_{DG}=0.03$ is, thus, less massive compared to the corresponding one in the disk with $f_{DG}=0.01$. Being so close to the inner edge of the disk, the planet originating at 1 AU reaches it before the disk dispersal, as all of the innermost planets for the different example simulations that we discussed above. 
 
 We have shown above that with a very low disk mass, the planets in these examples cannot grow efficiently. One wonders, whether increasing the dust-to-gas ratio can compensate for a lower disk mass. Indeed, as shown in Fig. \ref{Fig:Growthtracks_moreM0moredtg}, increasing the dust-to-gas ratio of the disk provides the necessary solid content for efficient core growth. We note that in this plot we compare two sets of simulations with $f_{DG}$=0.01 and $f_{DG}$=0.03 (in contrast to the standard case of Figs. \ref{Fig:Growthtracks_main} and \ref{Fig:Growthtracks} with $f_{DG}$=0.015). For an initial disk mass of 0.04 $M_{\odot}$ and the enhanced dust-to-gas ratio of $f_{DG}$=0.03 (purple colors), the planet starting at 10 AU reaches the pebble isolation mass within 1 Myr; therefore, it has enough time left to accrete gas efficiently. It becomes a Jupiter-mass gas giant, in contrast to the planet at the same starting position in the disk of the same mass with $f_{DG}$=0.01 (green colors). The same is true for a disk mass of 0.07 $M_{\odot}$ and the planet starting at 30 AU. While it failed to reach the pebble isolation mass with the lower dust-to-gas ratio, it reaches almost 3 $M_{J}$ with $f_{DG}$=0.03. 
 In Fig. \ref{Fig:Growthtracks_moret0moredtg}, we also test later embryo injection times with a higher dust-to-gas ratio and find a minimal difference in the final planetary masses. Even with the enhanced solid content, there is not enough time for the cores to grow efficiently if they did not do so with $f_{DG}$=0.01, given that most of the pebbles have already drifted inwards before the embryo is injected.

\begin{figure*}
\centering
\includegraphics[width=\textwidth]{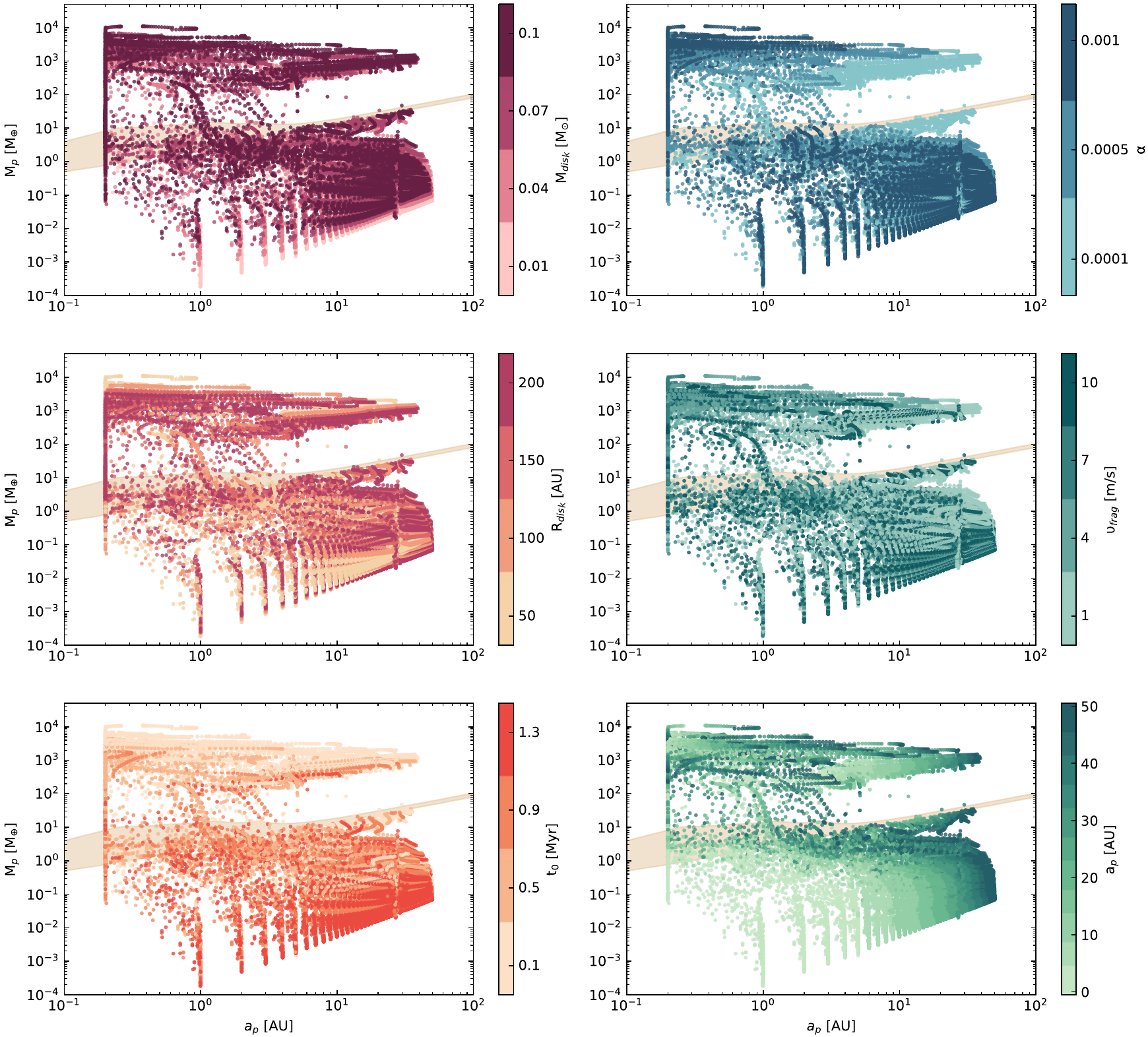}
\caption{Final planetary masses and final positions for all simulations with $f_{DG}$=0.015. The color-coding in each plot represents the different initial conditions tested for the corresponding parameter. The beige area shows the range of the initial pebble isolation masses, as calculated by the disk properties.}
\label{Fig:ap-Mp}
\end{figure*}

\begin{figure*}
\centering
\includegraphics[width=\textwidth]{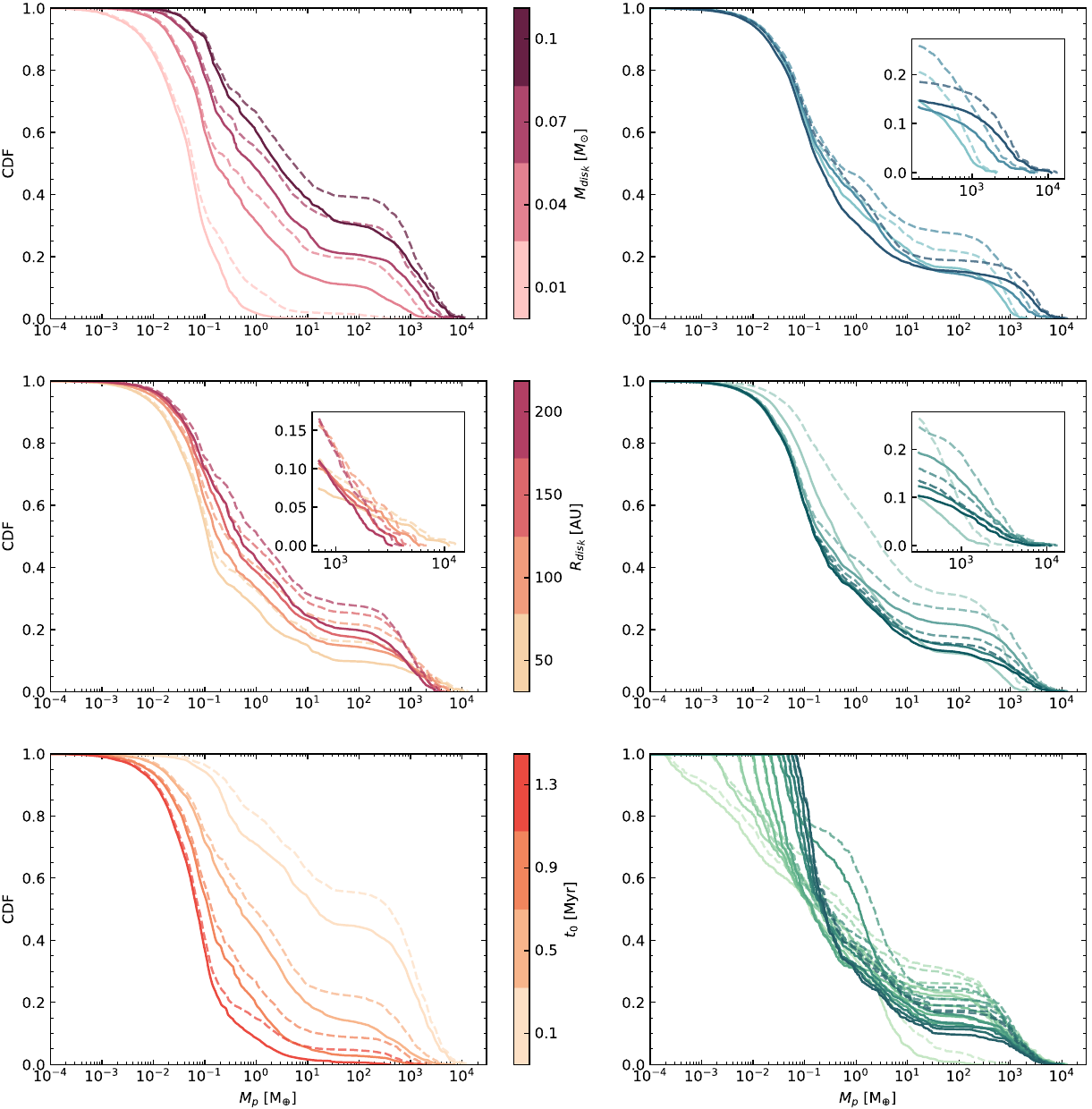}
\caption{Cumulative distribution functions (inversed) of final planetary masses, for different initial conditions. Each plot shows the effect of one parameter and the color-coding shows the different values for each parameter. The solid lines show simulations with $f_{DG}$=0.015, while the dashed lines show the ones with $f_{DG}$=0.03.}
\label{Fig:CDF-Mp}
\end{figure*}

\section{Favorable conditions for giant planet formation}
\label{Sec:Giant planet recipes}

We can now zoom out to the results of our whole sample of the parameter study. The parameters that we varied are presented in Table \ref{Tab:parameters} and all possible combinations have been tested, for two dust-to-gas ratio values ($f_{DG}$=0.015 and $f_{DG}$=0.03), yielding a total number of 76800 runs. In Fig. \ref{Fig:ap-Mp}, we present the final planetary masses as a function of final orbital distance, color-coding for the different initial parameters, only for the nominal $f_{DG}$=0.015. The beige region encloses the initial pebble isolation masses for all simulations, as calculated by each corresponding disk. 

The inner edge of the disk in our model is at 0.2 AU and this causes the pile-up of the planets in a vertical "line" at this position of the plots. The other vertical "lines" further out and near the bottom of the plots come from our choice of initial positions for the embryos. The gap, roughly above the maximum pebble isolation mass, is caused by the nature of gas accretion. If the conditions are such that a core can reach the pebble isolation mass, then envelope contraction most likely happens fast and runaway gas accretion begins. The pebble isolation mass generally increases with increasing orbital distance (Eq. \ref{Eq:pebiso}) and the gas accretion rates scale with the core mass. Therefore, the planets that reach the pebble isolation mass will have an increasing core mass as a function of orbital distance. This leads to this gap by gas accretion in Fig. \ref{Fig:ap-Mp} to become larger at larger distances.

 We also notice a more distinct diagonal "line" of points in Fig.\ref{Fig:ap-Mp} around 1 AU, which corresponds to embryos with an intermediate disk radius of 100 AU with otherwise beneficial initial parameters for core growth; high disk mass, low viscosity, and initial position of the embryo at 5 to 25 AU. In these cases, the fragmentation velocity is 10 m/s and this leads to larger dust grain sizes that drift at greater speeds. However, the embryos in these cases are injected at 0.5 Myr, so there are still enough pebbles left in the disk. The low accretion rates mean that migration outperforms growth and the type I inward migration of the planets stops when they reach the pebble isolation mass around 1 AU because of the enhanced surface density exterior to the water ice line. Then the disk is dispersed before they can accrete more gas and they remain at the level of sub-Neptune to Jupiter masses. 

Even though we vary a lot of the initial parameters in the simulations, we did not randomize the choices as in population synthesis simulations \citep[e.g.,][]{IdaLin2004,Mordasini+2009,Ndugu+2018,Mulders+2019,Emsenhuber+2021,Voelkel+2022}, noting that determining  distributions of the initial condition parameters still harbors several uncertainties. For example, it is still unknown what the initial mass distribution of protoplanetary disks is. The scope of this work is to investigate what initial conditions favor, from a theoretical point of view, the formation of giant planets, without comparing to the observed exoplanet populations and occurrence rates. For this reason, we have chosen our initial parameters within reasonable ranges constrained by observations, laboratory experiments, and theory, focusing more on the conditions which would be beneficial to giant planet formation.

We show in Fig. \ref{Fig:CDF-Mp} the (inverse) cumulative distribution functions (CDF) as a function of the final planetary masses, which describe the probability of finding a simulated planet of our sample above a certain final mass. 
In these plots, we also used color-coding to show the values tested of a specific parameter, for instance, the disk mass in the first one, including all of the possible combinations of the other parameters. The solid lines represent the simulations with $f_{DG}$=0.015 and the dashed lines show the simulations with $f_{DG}$=0.03. 

In the following discussion, we focus on the most favorable conditions to make a gas-giant planet. As such, we refer to any planet with a mass above 100 $M_{\oplus}$, and given that our disks extend only down to 0.2 AU, we discuss the analogs of warm and cold Jupiters. As mentioned above, we do not use randomized initial parameters and do not intend to directly compare with the occurrence rates of giant planets, so the presented fractions represent a measure of how "easy" it was to form a giant planet given the initial conditions compared to all of the planets formed in our simulations. However, we can conclude that if a specific parameter produces a much higher fraction than~10-20\% \citep[e.g.,][]{Cumming+2008,Mayor+2011,Wittenmyer+2016,Fernandes+2019} then this means that the initial condition that produced it cannot be the main condition that is true for the exoplanet population. Inversely, if a condition does not reproduce the occurrence rates, then it is probably also not the main condition leading to exoplanet formation. In the following sections, we discuss how each parameter influences planet formation, focusing on the formation of giant planets. 

\subsection{Disk mass}
\label{subsec:disk mass}
The total disk mass determines the total content of the building blocks that form planets. The formation of giant planets is not necessarily a two-step process (i.e., of pebble and then gas accretion) but regardless of when gas contraction happens and how the efficiency evolves, giant planets require the formation of a massive core. We expect that having large disk solid contents will be beneficial given that this will allow for higher pebble accretion rates. Once a sufficiently massive core has formed, a greater gas component can also contribute to efficient gas accretion that will then increase the total planetary mass. 

In the top-left plot of Fig. \ref{Fig:CDF-Mp}, we can see that a clear trend emerges with the increasing disk mass. An initial mass of 0.01 $M_{\odot}$ does not provide enough solid material for cores to grow and this makes it impossible for the planets to accrete gas and grow to Jupiter masses. Then, we find that gas giants with masses above 100 $M_{\oplus}$ make up 10\% of our sample with a disk mass of 0.04 $M_{\odot}$, 20\% with $M_{disk}$ = 0.07$M_{\odot}$, and 30\% with $M_{disk}$ = 0.1$M_{\odot}$. In the simulations with $f_{DG}$=0.03, we find that 10\% of our sample with a disk mass of 0.01 $M_{\odot}$ form a planet with a mass higher than 1 $M_{\oplus}$ but only around 1\% of the sample is made up of giants. 

With the increased dust-to-gas ratio, the initial dust mass increases, therefore the available content for core growth increases; then, if the cores are more massive,  the gas accretion rates will also increase, as discussed above. For example, the CDF with a disk mass of 0.04$M_{\odot}$ and $f_{DG}$=0.03 (dashed lines) is very similar to the one with $M_{disk}$ = 0.07$M_{\odot}$ and the nominal $f_{DG}$=0.015. Accordingly, the CDF with $M_{disk}$ = 0.07$M_{\odot}$ and $f_{DG}$=0.03 is almost the same as the one with $M_{disk}$ = 0.1$M_{\odot}$ and $f_{DG}$=0.015. 

\subsection{Disk radius}
\label{subsec:disk radius}
A large disk radius ensures a long-lasting pebble flux which benefits core growth via pebble accretion. We show in the middle-left plot of Fig. \ref{Fig:CDF-Mp} that the different initial disk radii lead to similar fractions of giant planets, decreasing from 20\% for a 200 AU disk to 10\% for a 50 AU disk. Interestingly, the trend is flipped for high planetary masses, as shown in the zoomed inset. The planetary mass at which this "flip" happens is strongly influenced by our choice of initial parameters for the whole set of simulations. Therefore, the critical mass cannot be quantified from this work, but we can conclude that if the conditions are favorable for a timely core growth to several Earth masses and subsequent gas accretion, then a smaller disk leads to more massive giants. The smaller disk radius means that more solid and gas material is in the vicinity of the planets and their masses can increase rapidly. Essentially, the smaller disk has a higher accretion rate due to the increased surface density.
 
The fraction of giants present is significantly enhanced if we increase the dust-to-gas ratio; specifically, we find that giants are more than 20\% of our sample with a disk size of 100 AU, which is higher than the largest disk size (200 AU) with half the dust-to-gas ratio and they are 35\% for the largest disk size (200 AU). There is again a flip for the highest planetary masses: the larger the disk, the lower the fraction of very massive giants. Additionally, above 1000 $M_{\oplus}$ the difference between the models with $f_{DG}$=0.03 and the ones with $f_{DG}$=0.015 gradually diminishes.

\subsection{Starting time of the embryo}
As expected, we find that the injection time of the embryo plays a decisive role in the growth of a giant planet. When the planetary embryo starts growing 0.1 Myr after the modeled disk starts evolving, then the fraction of giants produced is almost 45\% (left, bottom plot in Fig. \ref{Fig:CDF-Mp}) and 50\% with $f_{DG}$=0.03. In contrast, injecting the embryo as late as 1.3 Myr leaves us with almost no giants. In the latter case, there is simply not enough material to build the core fast because the pebbles have mostly drifted inwards. Even if the core reaches pebble isolation mass before the disk dispersion, it will do so very late, leaving insufficient time to accrete large amounts of gas. This can be marginally improved by increasing the dust-to-gas ratio. In this case, we find <5\% of giants in our sample for a starting time of 1.3 Myr. 

The few giants that are formed with this late injection time had very beneficial initial parameters otherwise, as we see in the few dots above 100 $M_{\oplus}$ in Fig. \ref{Fig:ap-Mp}.
Specifically, those included high disk mass, large disk radius, low viscosity and fragmentation velocity (Sect. \ref{subsec:frag vel}), and starting locations at 2 to 19 AU. Reducing the injection time to 0.9 Myr is still not very helpful, as it only produces around 2.5\% giants and the fraction is below 15\% even for an injection time of 0.5 Myr. Nevertheless, the corresponding fractions with $f_{DG}=0.03$ are 10\% and approximately 23\%.

We should also mention that the implications of the starting time for the embryo are heavily influenced by the dust fragmentation velocity. A low fragmentation velocity results in small pebbles and, thus, they are kept in the disk longer, while a disk with a larger fragmentation velocity will get depleted of its pebbles too fast. This can be seen, for example, in Fig. \ref{Fig:ap-Mp} (left, bottom and right, middle plots), where the few giants that are formed when the embryo is injected at 1.3 Myr correspond to a disk with $u_{frag}=1$ m/s, which retained the pebbles long enough for the planets to reach the pebble isolation before the gas dissipation, so that they still had time to accrete gas.

\subsection{Disk viscosity}
\label{subsec:viscosity}
A higher $\alpha$-viscosity parameter means that the viscous heating is higher, therefore the temperature of the disk and its aspect ratio increase, and this leads to higher pebble isolation masses in the inner disk. The increased viscosity also means increased destructive collisions between dust grains and this decreases the available pebble-sized material, which, in turn, leads to decreased pebble accretion rates. In total, this results in a less efficient formation of planets of a few Earth masses.

On the other hand, gas accretion benefits from the cores that could be formed from the higher pebble isolation mass. Hence, viscosity plays a different role for low-mass and high-mass planets, similar to the disk radius (top-right plot in Fig. \ref{Fig:CDF-Mp}). We find that very high-mass giants can be formed with high viscosity and for this population, the higher the viscosity, the more massive the giants become. However, the key in   this situation is early formation  (see also Fig. \ref{Fig:ap-Mp}), so that the core can take advantage of as much solid material as possible before it is lost due to radial drift, and the fact that at high viscosity the gas accretion rate is higher. Even when the horseshoe region has been emptied, a higher viscosity will provide more material to the planet and it will thus become more massive. 

Higher viscosity also leads, in general, to faster migration rates because it delays the gap opening to higher planetary masses, so that the planets remain in type-I migration longer and migrate inwards further before the slow type-II migration sets in.
At the same time, at high viscosity, the entropy-driven corotation torque can operate, which would actually slow down type-I migration. However, type II migration is directly dependent on the viscosity of the disk, so a lower $\alpha$ value will slow down migration for the planets that have opened a gap in the disk and prevent them from migrating to the inner edge of the disk.
For this reason, we find ( see Fig. \ref{Fig:ap-Mp}) that high-mass gas giants have a decreasing final position for increasing $\alpha$-viscosity values. 

The overall difference in the CDFs between various $\alpha$-viscosity values is small, with all of them leading to similar fractions of gas giants in our sample around 15-18\%. In contrast, the increased dust-to-gas ratio models lead to more distinct CDFs that are also flipped and the highest $\alpha$-viscosity leads to the lowest fractions of planets up to a few hundred $M_{\oplus}$. This was also the case with $f_{DG}$=0.015 but up to lower planetary masses because of the less efficient core growth and the lower pebble isolation masses.
The fraction of giants with $\alpha=10^{-3}$ and $f_{DG}=0.03$ is higher than for all $\alpha$-viscosities with the nominal dust-to-gas ratio. Then it is around 25\% for $\alpha=10^{-4}$, while the highest fraction, namely, of 35\%, comes from the disks with the intermediate $\alpha=5\times10^{-4}$, which contributes to maintaining a balance between the pebble and gas accretion rates. 

\subsection{Dust fragmentation velocity}
\label{subsec:frag vel}
The fragmentation velocity defines the threshold relative velocity at which collisions between dust grains lead to destruction into smaller aggregates. The higher this threshold, the larger the dust grains become. This is not only directly beneficial to pebble accretion, but it also increases the radial drift velocities, which can then increase the pebble flux towards the growing core. Nevertheless, this can actually hinder core growth if the planetary embryo is injected later into the disk and a significant fraction of the dust has already drifted inward to the embryo location.

The most beneficial fragmentation velocity out of those tested is 4 m/s with a fraction of gas giants slightly more than 20\% of the whole sample, while the 7m/s threshold results in a 15\% giant planet fraction (middle-right plot in Fig. \ref{Fig:CDF-Mp}). These intermediate values offer the necessary pebble sizes for maximized accretion and, at the same time, their radial drift velocities provide an optimal pebble flux, which can also last a long time. In contrast, the highest value of 10 m/s and the lowest of 1 m/s both yield a fraction of approximately 10\%. 

When the fragmentation velocity is low, the dust particles are kept small, hence, they are mostly well coupled to the gas and so, pebble accretion is delayed. It takes longer to reach the pebble isolation mass and, consequently, this leaves less time for the accretion of a massive envelope. At the same time, the smaller pebbles mean lower radial drift velocities, so the pebble flux can be maintained longer. When the fragmentation velocity is high, then large particles can form,  allowing for improved accretion from the embryo. However, the high Stokes numbers also lead to high radial drift velocities, which require an early formation of the core so that they are not lost before accretion onto the embryo can start.

Interestingly, with $f_{DG}=0.03$, the most favorable fragmentation velocity is 1 m/s, with the fraction of giants exceeding 30\% of our sample. This can be linked to the decreased drift velocities, in combination with the higher solid content, that compensates efficiently for the smaller pebble sizes, given that the pebble accretion rates have a stronger dependence on the surface density than the (Stokes numbers of the) pebble sizes. The fractions drop significantly beyond 1000 $M_{\oplus}$ and become again the lowest compared to higher fragmentation velocities.  

\subsection{Starting location of the embryo}
\label{subsec:position}
In the right, bottom plot of Fig. \ref{Fig:CDF-Mp}, we only show some of the starting positions for better readability. At 1 AU, almost no embryos grow to become gas giants because they migrate to the inner edge of the disk, where we stop the accretion onto the planet. The highest fraction of giants can be found with embryos that start at 5 AU and then it decreases up to 25 AU. The embryos that originate at 30 AU then have higher fractions of giants compared to those interior to this orbital distance, but injecting the embryo further out results in a decrease to the fraction of giants again because the further out the embryo is located, the closer it is to the outer edge of the disk (especially for the very small 50 AU disks), where the solids' surface density is significantly lower and pebble accretion needs to surmount radial drift. 

The fractions of very massive giants ($M_p\geq$ 1000 $M_{\oplus}$), are very similar for all starting locations, except for the ones starting their growth at 1 AU. Reaching such high mass is a product of a combination of beneficial conditions and an absence of adverse ones, such as high disk mass (0.1 $M_{\odot}$) and early injection of the embryo (0.1 Myr), viscosity dictated by $\alpha\geq5\times10^{-4}$, and fragmentation velocities higher than 1 m/s, but it dependents weakly on where the embryo originates. 

With the increased dust-to-gas ratio, we find $\sim$5\% of giants form even when the embryo originates at 1 AU because the enhanced solid content aids the core growth before the planet migrates to the inner edge and stops accreting. In general, the enhanced core growth with $f_{DG}$=0.03 leads to less inward migration. The highest fraction of giants corresponds again to starting their growth at 5 AU because of the high solids surface density that is additionally enhanced at the water ice line and the fact that the pebble flux exterior to them can be sustained longer, compared to starting locations further out. The location of the water ice line is close to 1 AU, but it depends on the disk properties that define viscous heating, specifically, the disk mass that sets the gas surface density, the viscosity, and the dust-to-gas ratio; therefore embryos originating at 1 AU could also benefit from the higher solids surface density at the water ice line in some of the cases. However, they most likely migrate to the inner edge of the disk, where we stop their growth, unlike the planets that start growing at 5 AU. Additionally, the planets originating at 5 AU can migrate outwards (depending on the disk parameters) due to the heating torque and this also prevents them from migrating to the inner edge (very early). 

\subsection{Dust-to-gas ratio}
A higher dust-to-gas ratio means that the solid component of the disk is increased and that the initial pebble isolation mass is higher because of an increase in the viscous heating component of the temperature (Eq. B.3 in \citealt{SchneiderBitsch2021a}). It should be noted, however, that the pebble isolation mass is also influenced by the mean molecular weight that is evolving in time and space as the inward drifting dust evaporates at the icelines and the gas vapor content gets enhanced (Eq. E.13 in \citealt{SchneiderBitsch2021a}) and this can lead to a lower pebble isolation mass in the inner disk. 

Here, we compare $f_{DG}$=0.03 to the nominal value of $f_{DG}$=0.015. We see in the dashed lines of Fig. \ref{Fig:CDF-Mp} (for all different parameters) that the small planets, around 1 $M_{\oplus}$ and below are not significantly influenced by an increase in the dust-to-gas ratio because, in most of the cases, the conditions remain limiting for planet growth, independently of the dust content, and the fact that the pebble isolation mass can be higher makes it harder to reach this mass. Yet, overall and as expected, the higher solid content leads to higher planetary masses and thus higher fractions of giants. 

However, if the conditions are not favorable for giant planet formation, then the higher dust-to-gas ratio increases the fractions but cannot compensate efficiently enough. For example, when the embryo is injected late in the disk (bottom-left plot of Fig. \ref{Fig:CDF-Mp}), we find in this case to account for less than 5\% of giants compared to almost zero with the nominal dust-to-gas ratio. In this case, a significant amount of pebbles is still lost due to radial drift before the core starts to accrete material. Similarly, even though we find that almost no giants form when the embryo originates at 1 AU, the fraction is also around 5\% with double the dust-to-gas ratio because the enhanced pebble accretion helps in reaching the pebble isolation mass sooner and slows down their migration. On the contrary, we find that the lowest disk mass we tested ($M_{disk}$=0.01 $M_{\odot}$) is very limiting and even for the increased amount of solids, the pebble flux is too low for planet formation in general and prohibits the formation of giants. 

\section{Discussion}
\label{Sec:Discussion}

Global models have been increasingly emerging in an effort to self-consistently link the dust and gas evolution with planet formation. \citealt{Bitsch+2019} investigated the conditions around the formation of gas giants, located outside 1 AU. They used N-body simulations, while varying the pebble flux and the viscosity in relation to migration and accretion and keeping the viscosity of the disk fixed to 0.0054. In the accompanying work, \citealt{Lambrechts+2019} used the pebble flux as a free parameter. Their findings are in line with our findings in the present work but our goal is to generalize the conclusions about the favorable conditions for giant planet formation and to link it directly to the conditions dictated by the interplay with the protoplanetary disk. 

\citealt{Lambrechts+2019} find that if the total dust mass is less than around 110 $M_{\oplus}$, then the masses of the formed planets are below 0.1 $M_{\oplus}$, even with the inclusion of growth through embryo collisions after the gas dissipation. Here, our 0.01 $M_{\odot}$ disks correspond to a total $M_{dust} \approx$ 50 $M_{\oplus}$ for the nominal $f_{DG}$=1.5\% and these produce almost no planets with masses above 1 $M_{\oplus}$ (Fig. \ref{Fig:CDF-Mp}). Growth is highly dependent on the Stokes number, which can be slightly different between our models, given some different initial conditions. In total, however, it is not surprising that even with $f_{DG}$=3\% we find only $\leq$ 2.5\% of the final planetary masses of our sample are above 10 $M_{\oplus}$. 

Even though we focus here on gas giant formation (like Jupiter and Saturn), we have shown that ice giants (like Neptune and Uranus) can also be formed with our model. According to the model proposed in \citealt{Lambrechts+2014}, ice giants are planets in wide orbits that do not grow enough to reach their pebble isolation masses before the gas is lost, so the heating from the continued pebble accretion prevents them from contracting a massive gaseous envelope. In our example cases, we show in the middle-left plot of Fig. \ref{Fig:Growthtracks} an embryo that originates at 30 AU, accretes pebbles slowly because of the limited supply this close to the outer edge of the disk, and by the end of the disk's lifetime, it is just above 10 $M_{\oplus}$ but has not reached the pebble isolation mass yet, which means that the planet is still hot and might not go into envelope contraction. Similarly, in Figs. \ref{Fig:Growthtracks_moreM0moredtg} and \ref{Fig:Growthtracks_moret0moredtg}, some of the outermost embryos have final masses around 10 $M_{\oplus}$ and in Fig. \ref{Fig:ap-Mp}, we show several planets in wide orbits (>10 AU) with masses of 10-50 $M_{\oplus}$ and just around their pebble isolation mass, which means that they have not undergone runaway gas accretion. Interestingly, all of these potential ice giants are formed in disks with $\alpha=0.0001$ (top-left plot of Fig. \ref{Fig:ap-Mp}). 

In this work, our model does not yet include planetesimal accretion, however, planetary cores undergoing formation via pure planetesimal accretion \citep[e.g.,][]{IdaLin2004,Mordasini+2009,Mordasini+2012,Emsenhuber+2021} face many difficulties, especially with giant planet formation \citep{Drazkowska+2022arXiv}. This is mainly due to the long growth timescales that either compete with the disk lifetimes \citep{Pollack+1996,Fortier+2013,Guilera+2014} or are overpowered by migration \citep{Voelkel+2020}.
\citealt{Bruegger+2020} presented a comparison between pebble and planetesimal accretion in population synthesis models, concluding that pebble accretion is less efficient in giant planet formation. Their study is limited to a disk with $M_{disk}$=0.017 $M_{\odot}$, $\alpha$=0.002, $f_{DG}$=0.01, and $R_{disk}$=30 AU. As we have shown in the present work, this combination of initial conditions is unfavorable for giant planet formation, given that all of them prevent efficient pebble accretion. 

\citealt{Voelkel+2022} concluded that even though efficient growth can happen via pebble accretion, it might actually be a destructive mechanism for inner disk embryos that form early on because of fast and extended migration. However, these authors also stress that the choice of initial disk parameters is limited; specifically, they used $M_{disk}$=0.1 $M_{\odot}$, $\alpha$=0.0003, $f_{DG}$=0.0134, $R_{disk}$=20 AU, and $u_{frag}$=2 m/s. In the present work, we find that these specific parameters are not necessarily inefficient for planet growth, in line with their conclusions, even though this combination is probably not one of the most favorable. Several planets reach indeed the inner edge of the disk but, as we show here, this mainly happens  to the embryos that originated at 1 AU, it is, thus, expected because of the small disk radius they are testing and it is not a common phenomenon in planet formation models. In general, we show here that the pebble accretion framework can be very efficient at forming giant planets (and broadly planets), especially when at least one of the initial conditions is beneficial. 

In the present work, as well as in several pebble accretion models, the initial planetary embryos are assumed to be already formed and their injection locations are handpicked (what we do here) or chosen from a distribution in a randomized way. This does leave space for questions on how these embryos form and whether some disk locations are favorable or forbidden. \citealt{Voelkel+2020} linked the dust and gas distribution to the formation of planetesimals and subsequently to the spatial distribution of planetary embryos and show that the location is crucial, with the innermost regions of the disk being more favorable due to the higher concentration of solid content. 

At the same time, other studies have discussed that certain locations in the disk can be the birthing locations of planetary embryos, such as the water ice line \citep{SchoonenbergOrmel2017,DrazkowskaAlibert2017,Mueller+2021} or the silicate sublimation line \citep{Izidoro+2021,Morbidelli+2022}. Our choice of a wide range of initial locations serves the scope of this paper to connect the starting position of an embryo with the final position and mass of a formed planet, regardless of whether some of these locations are more probable than others. Additionally, we conclude that forming giant planets is not severely limited by the initial location of the embryo but some locations are more favorable (Fig \ref{Fig:CDF-Mp}), for example around 5 AU and around 25-30 AU (depending also on the rest of the disk parameters).

\citealt{Coleman2021} used a global model that includes both pebble and planetesimal accretion, along with planetesimal formation, to study the embryo formation and determine their masses, sizes, and spatial distributions in the disk. In the combined (pebble \& planetesimal) accretion scenario, they find that planetesimal accretion is mainly effective around the iceline and a few AU exterior to it. They also find that including planetesimals can actually aid in pebble accretion, despite the fact that planetesimal formation will end up using some of the available mass in solids \citep{Voelkel+2021}. Specifically, planetesimal accretion can help the planetary embryos become more massive and reach the transition mass sooner, thus enhancing the efficiency of pebble accretion. 

The dust-to-gas ratio can be used as a proxy for the metallicity of the disk through
\be [Fe/H]=log_{10}\left(\frac{f_{DG}}{f_{DG,\odot}}\right)~.\ee
Setting our nominal simulations at [Fe / H] = 0 leads to [Fe / H] =$\log_{10}{(3/1.5)}\approx0.3$ dex for the simulations with $f_{DG}=0.03$.
Previous studies have established a positive correlation between metallicity and giant planet occurrence rate \citep{Santos+2004,FischerValenti2005,Johnson+2010,Fulton+2021}. In the pebble accretion framework, the core mass is heavily influenced by the initial dust-to-gas ratio \citep{Lambrechts+2014} and the heavier cores will contract faster and move on to rapid gas accretion sooner \citep{Ikoma+2000}. 

We find here, unsurprisingly, that a higher dust-to-gas ratio leads, in general, to more massive planets (Fig. \ref{Fig:CDF-Mp}). However, another factor could be preventing giant planet formation. As an example, we show in Fig. \ref{Fig:Growthtracks_moreM0moredtg} that the higher dust-to-gas ratio could provide enough material so that migration slows down and the pebble isolation mass can be reached sooner to give way to efficient gas accretion. However, in Fig. \ref{Fig:Growthtracks_moreM0moredtg}, we do not include the lowest disk masses we tested (0.01 $M_{\odot}$) because in this case, the higher dust-to-gas ratio cannot compensate efficiently for the low solid content -- and, again, almost no giants are formed. Additionally, we see in Fig. \ref{Fig:Growthtracks_moret0moredtg} that the early formation is more important than the high dust-to-gas ratio due to the radial drift that depletes the disk from the solids even before the planet starts growing. In the general picture (bottom-left plot in Fig. \ref{Fig:CDF-Mp}), there is only a minimal increase in the fraction of giants produced with $f_{DG}$=0.03 and late injection times. 

Interestingly, though, we see in Fig. \ref{Fig:CDF-Mp} that the CDF with a disk mass of 0.04 $M_{\odot}$ and $f_{DG}$=0.03 is very similar to the one with a disk mass of 0.07 $M_{\odot}$ and $f_{DG}$=0.015 and the CDF of the disks with $M_{disk}$ = 0.07 $M_{\odot}$, $f_{DG}$=0.03 are almost the same as that of the disks with $M_{disk}$ = 0.1 $M_{\odot}$, $f_{DG}$=0.015. It is thus very clear that further constraints (from observations and theory) on the initial disk mass fraction are needed as input for planet formation simulations.

One could argue that our choice of initial disk masses is on the higher side (e.g., as seen in Fig. 6 in \citealt{Manara+2022arXiv}) and is thus expected to form more giant planets, but not necessarily in line with the observed disk masses. Nevertheless, we note that, on the one hand, the mass estimates come from observations around mm wavelength and, on the other hand, the disk mass is evolving over time and as the dust drifts radially, a significant percentage of it gets accreted, either by forming planets or by the host star. We will expand on this dust mass evolution in future work, but we can already conclude that this process will naturally decrease the dust mass over time. Hence, the remaining disk mass of our models closer to the dissipation of the gaseous disk would also populate the lowest disk masses presented in \citealt{Manara+2022arXiv}.

In this work, we have tried to use the existing observational constraints in order to determine the ranges of our tested parameters and we have discussed some of our assumptions. Even if we improved our assumptions or added new physics to our models, we would not expect the general trends to be significantly altered; rather, any changes would be seen in the specific fractions for giant planets (or other types of planets in general). We want to note that significant constraints can be made from our findings even for initial parameters that overproduce giants (e.g., $M_{disk}$=0.1 $M_{\odot}$ or $t_0$=0.1 Myr) or struggle to produce any planet at all (e.g., $M_{disk}$=0.01 $M_{\odot}$). The latter, for instance, places a constraint on the masses of the disks that we would expect to host a planet unless other mechanisms can increase the planetary masses. Such a mechanism could be growth by collisions \citep{KominamiIda2004,OgiharaIda2009,Cossou+2014,Izidoro+2017,Ogihara+2018,Izidoro+2019}, however, we would not expect the final masses to be enhanced by more than a few $M_{\oplus}$ \citep{Lambrechts+2019}. 

We have also shown that even with a higher dust-to-gas ratio, a very small fraction of planets with masses above 1 $M_{\oplus}$ do form, therefore, we would expect that a limiting disk mass exists, especially for higher-mass planets. Similarly, there are also cases of a combination of conditions making it harder for giants to form. For example, the viscosity of the disk or the fragmentation velocity do not seem to place significant constraints (e.g., comparing the giant planet fractions in Fig. \ref{Fig:CDF-Mp} for different values of these parameters), however, a combination of high viscosity ($\alpha$=0.001) and low fragmentation velocity ($u_f$=1 m/s) leads to very small dust particles and, thus, low drift velocities and pebble accretion rates, thus hindering the growth of giant planets.

We cannot point to a specific parameter being a determining initial condition for gas giant planet formation, but it is rather a combination of beneficial factors that are in play. If we wanted to reproduce, for instance, a specific system or star-forming region, we could use an appropriate "mixture" of the initial conditions that would reproduce the observed conditions accordingly. For these reasons, we would expect the diversity of the exoplanetary systems to come from an intrinsic diversity in their natal protoplanetary disks.  The most important conclusion in this work is that we need to understand and constrain the underlying disk population better in order to understand and constrain planet formation.

\section{Summary and conclusions}
\label{Sec:Summary}
In this paper, we connect the initial conditions in a protoplanetary disk to the formation efficiency of giant planets. To do so, we performed 1D semi-analytical numerical simulations of planet formation via pebble \citep{JohansenLambrechts2017} and gas accretion \citep{Ndugu+2021} in a viscously evolving protoplanetary disk using \texttt{chemcomp} \citep{SchneiderBitsch2021a}. Our model includes a two-population approach for dust growth and drift \citep{Birnstiel+2012}, evaporation at the icelines\footnote{Multiple chemical species are considered in the disk, however, we do not discuss the composition of the planets here and they only affect our work by the spikes of the surface density around the icelines.}, type-I \citep{Paardekooper+2011} and type-II migration, the effects of thermal \citep{Masset2017} and dynamical torques \citep{Paardekooper2014}, and gap opening, as described in \citealt{Ndugu+2021}. We assume that planetary embryos have already formed in the disk, with masses equal to the transition mass \citep[Eq. \ref{Eq:M_t},][]{JohansenLambrechts2017}, thereby dependent on the local conditions and at which the embryos accrete pebbles in the (relevant to larger protoplanets) Hill (shear) regime.

We performed a parameter study, testing different values for the mass of the disk, the radius of the disk, the time when the embryo is injected, the $\alpha$-viscosity parameter, the dust fragmentation velocity, the location where the embryo is injected, and the dust-to-gas ratio of the disk (Table \ref{Tab:parameters}). 
We summarize our findings for each parameter as follows:
\begin{itemize}
\item A high disk mass is very important in the formation of a gas giant because they need a massive core that has reached the pebble isolation mass and can start accreting gas efficiently. Of course, the high total disk mass can also provide more gas to the planet during gas accretion. We find that with a disk mass of 0.01 $M_{disk}$, no giant planets can form, even in those disks with a higher dust-to-gas ratio.
\item A large disk radius ensures a long-lasting pebble flux that aids in core formation. However, we find that if the conditions are favorable for a giant planet to form in the first place, then the smaller the disk, the more massive the planet can get. Given that the total mass is the same, the increased surface density leads to enhanced pebble and gas accretion rates.
\item The time when we inject the embryo plays a decisive role because there is competition with the radial drift of the pebbles. When the embryo is injected, enough mass in the form of pebbles needs to remain in the disk and this is also determined by the fragmentation velocity that sets the pebble sizes and consequently the radial drift velocities. We find that when the embryo starts to grow at 1.3 Myr, it is very improbable that it can grow beyond 10 $M_{\oplus}$; thus, very few planets reach the pebble isolation mass and accrete any gas at all. A higher dust-to-gas ratio can make this late starting time less restricting, however, the fraction of giants that formed is still low, giving further indication that planet formation might start early in disks \citep[e.g.,][]{Segura-Cox+2020}. 
\item Viscosity, and thus viscous heating, directly influences the temperature of the disk, which changes its aspect ratio and, consequently, the pebble isolation mass. An increase in the viscosity also increases the dust collisions, so it leads to smaller pebbles that are better coupled to the gas and less easily accreted by a growing planet. However, a high viscosity means higher pebble isolation masses and if the conditions are otherwise favorable for the core to reach the pebble isolation mass, then gas accretion is enhanced by the more massive core. At the same time, the high viscosity replenishes the horseshoe region with new gas faster, further enhancing gas accretion and increasing the envelope mass. The fact that there is a trade-off between core growth and gas accretion for different viscosities leads to small differences between the gas giant fractions from different $\alpha$ parameters. 
\item The fragmentation velocity sets the maximum grain size, thus determining the pebble accretion rates and the radial drift velocities. We find that the most beneficial value for giant planet formation is an intermediate one compared to results from laboratory experiments (namely 4 m/s); however, with double the dust-to-gas ratio, we find that the most favorable threshold velocity is 1 m/s because of the increased pebble surface density. In general, the pebble accretion rates are maximized with a combination of optimal pebble sizes and surface density, so the most favorable fragmentation velocity is strongly influenced by the available amount of solids.  
\item We find that the fractions of giants peak close to 5 AU and 30 AU, but, in general, giant planet formation is not strongly dependent on the starting location of the embryo. Injecting the embryos very close to the star increases the chances that they migrate to the inner edge well before the gas dissipation. In contrast, injecting them very close to the outer edge, where the pebble flux is reduced, significantly increases the growth timescales. 
\item The fractions of giants generally increase with increasing dust-to-gas ratio because the solid content that is used for core growth is higher, so the pebble accretion rates are higher and the pebble isolation mass can be reached sooner, giving way to efficient gas accretion and increasing the chances that the planet can accrete a massive envelope before the end of the lifetime of the gas disk. However, the improvement is marginal if the rest of the initial conditions are unfavorable for giant planet formation, as is the case, for example, when the disk mass is very low (0.01 $M_{\odot})$.

\end{itemize}

Even if one of the more adverse conditions exists in the protoplanetary disk, when the rest of the conditions are favorable or at least not adverse, then it is still possible for a giant planet to form. In general, giant planet formation is dictated by a combination of advantageous conditions and we cannot simplify this process by pointing to a single defining parameter. 

Even though we did not choose our initial conditions in a randomized way and, thus, we cannot directly compare with the occurrence rates of the exoplanet systems, we suggest that if one of the initial conditions tested in this work overproduces or underproduces giants (or planets in general) compared to the estimated occurrence rates of exoplanets, then the fraction of protoplanetary disks with these conditions should be small. Most importantly, we conclude that the diversity of the exoplanetary systems is directly linked to the diversity of their natal protoplanetary disk. Therefore, we need to examine and constrain the disk population better through observations and obtain realistic initial distributions for the initial conditions to understand and constrain planet formation.

\begin{acknowledgements}

B.B. and S.S. thank the European Research Council (ERC Starting Grant 757448-PAMDORA) for their financial support. S.S. is a Fellow of the International Max Planck Research School for Astronomy and Cosmic Physics at the University of Heidelberg (IMPRS-HD).                                 
\end{acknowledgements}

\bibliographystyle{aa}
\bibliography{PaperS3}

\end{document}